\documentclass[prd,twocolumn,groupedaddress,noshowpacs,nofootinbib]{revtex4}
\usepackage{graphicx}
\usepackage{dcolumn}
\usepackage{amssymb}
\usepackage{mathrsfs}
\usepackage{amsmath}
\usepackage{epsfig}
\usepackage[dvips]{color}
\usepackage{hhline}

\begin{document}

\title{Split Higgs triplet}

\author{Pei-Hong Gu}

\email{phgu@seu.edu.cn}

\affiliation{School of Physics, Jiulonghu Campus, Southeast University, Nanjing 211189, China}

\begin{abstract}

We find if a Higgs triplet with hypercharge has a special dimension-6 operator with the standard model Higgs doublet, i.e. a square of the trilinear triplet-doublet coupling, its scalar(pseudo-scalar) component can obtain a small quadratic term while its pseudo-scalar(scalar) and charged-scalar components can hold their masses heavy enough. Such split Higgs triplet can spontaneously develop a small vacuum expectation value to realize a Majorana neutrino mass generation without causing any high-dimensional lepton number violations including the well-known Weinberg dimension-5 operator. Alternatively it can mediate a non-standard neutrino self-interaction motivated by resolving the tension in Hubble constant measurements. This effective theory with rich observable phenomena can be induced by the Georgi-Machacek Higgs triplets at tree level or some dark matter fields at one-loop order.

\end{abstract}


\maketitle

\section{Introduction}

It is well known that a Higgs triplet can obtain an induced vacuum expectation value (VEV) through its trilinear coupling with the standard model (SM) Higgs doublet \cite{mw1980,sv1980,cl1980,lsw1981,ms1981}. This induced VEV makes the electroweak $\rho$ parameter deviate from unit so that it should be smaller than a few GeV to satisfy the experimental limits \cite{pdg2018}. In order to relax the $\rho$-parameter constraint, we can simultaneously introduce two Higgs triplets, i.e. one carries a nonzero hypercharge and the other is real without hypercharge. The $\rho$ parameter then can be held at unit even if the two induced VEVs are as large as a few tens GeV \cite{cce2019}. This is the so-called Georgi-Machacek model \cite{gm1985,sg1985}.

As a scalar, the Higgs triplet in principle can be expected to spontaneously break certain symmetry. Unfortunately, such spontaneous Higgs triplets contain some light components so that their exist can not be consistent with the SM precision measurements. For example, the neutral component of the spontaneous Higgs triplet with hypercharge will give an unacceptable invisible decay width of the $Z$ boson. Similarly, the neutral and charged components of the spontaneous Higgs triplet without hypercharge will significantly enlarge the decay width of the $Z$ and $W^{\pm}_{}$ bosons. Therefore, people conclude that the Higgs triplets for spontaneous symmetry breaking should have been ruled out experimentally.

On the other hand, the discovery of neutrino oscillations has indicated that three flavors of neutrinos should be massive and mixed \cite{pdg2018}. Moreover, the neutrinos should be extremely light to fulfil the cosmological observations \cite{pdg2018}. Currently, the best explanation for the tiny but nonzero neutrino masses seems to be the famous seesaw \cite{minkowski1977,yanagida1979,grs1979,ms1980} mechanism. In the tree-level \cite{minkowski1977,yanagida1979,grs1979,ms1980,mw1980,sv1980,cl1980,lsw1981,ms1981,flhj1989,ma1998} or radiative \cite{ma2006} seesaw models for generating a Majorana neutrino mass term, some heavy particles such as the type-II seesaw Higgs triplet with hypercharge can always mediate a lepton-number-violating Weinberg dimension-5 operator \cite{weinberg1979}. Alternatively, we can consider some dilepton, leptoquark or diquark scalars \cite{zee1986,babu1988,bl2001}. By integrating out these dileptons, leptoquarks or diquarks, we can obtain some high-dimensional lepton number violations and then can have the Majorana neutrino masses in association with the SM Yukawa couplings. For example, the singly-charged and doubly-charged dileptons in the Zee-Babu model \cite{zee1986,babu1988} can mediate a dimension-9 lepton-number-violating operator. To the best of our knowledge, the Majorana neutrino mass generation must be accompanied by the appearance of certain high-dimensional lepton-number-violating operators.

Furthermore, the discrepancy between the Hubble constant measurements from the early and late Universe observations has grown in significance to $4-6\, \sigma$ over several years and has posed a serious challenge to the standard $\Lambda$CDM cosmology. The low-redshift measurements of the matter density fluctuation amplitude on 8 Mpc scales, $\sigma_8^{}$, also appear to be lower than the $\Lambda$CDM prediction. The significant "Hubble tension" and the milder "$\sigma_8^{}$ tension" can be simultaneously resolved by the neutrinos with non-standard self-interactions \cite{cs2017,ah2017,lckp2017,otrw2017,kcd2019,fln2019,bkkm2019}. However, the viable realization of such non-standard self-interacting neutrinos was found to be highly challenging \cite{bkkm2019}.

In this paper, we shall present a novel idea of split Higgs triplet. Specifically, after the SM Higgs doublet drives the electroweak symmetry breaking, the scalar(pseudo-scalar) component of a Higgs triplet with hypercharge can obtain a small quadratic term while the pseudo-scalar(scalar) and charged-scalar components can hold their masses heavy enough to fulfil all experimental constraints, thanks to a special dimension-6 operator between this Higgs triplet and the SM Higgs doublet, i.e. a square of the trilinear triplet-doublet coupling. Such split Higgs triplet can spontaneously develop a small VEV to realize a Majorana neutrino mass generation without causing any high-dimensional lepton number violations including the well-known Weinberg dimension-5 operator. Alternatively, it can mediate a non-standard neutrino self-interaction to resolve the Hubble tension. This effective theory can be achieved at tree level or one-loop order in various renormalizable models with the Georgi-Machacek Higgs triplets \cite{gm1985,sg1985} or some dark matter fields \cite{ma2006,bhr2006,sz1985,cfs2006,lg2017,gh2019}.

\section{Split Higgs triplet}

The SM Higgs doublet and the Higgs triplet with hypercharge are denoted by
\begin{eqnarray}
\begin{array}{c}\phi(1,2,-\frac{1}{2})\end{array}\!&=&\left[\begin{array}{c}
\phi^{0}_{}\\
[1mm]
\phi^{-}_{}
\end{array}\right]\,,\nonumber\\
\Delta(1,3,+1)&=&\left[\begin{array}{cc}
\frac{1}{\sqrt{2}}\delta^{+}_{}& \delta^{++}_{}\\
[2mm]
\delta^0_{}\equiv \frac{1}{\sqrt{2}}\left(\delta^{0}_{R} + i \delta^{0}_{I}\right)& - \frac{1}{\sqrt{2}}\delta^{+}_{}
\end{array}\right]\,.
 \end{eqnarray}
Here and thereafter the brackets following the fields describe the transformations under the $SU(3)_c^{} \times SU(2)^{}_{L}\times U(1)_Y^{}$ gauge groups. Their full scalar potential at renormalizable level is expected to be 
\begin{eqnarray}
\label{potential}
V_{\phi\Delta}^{}&=&\mu_\phi^2 \phi^\dagger_{} \phi +\lambda_\phi^{}\left(\phi^\dagger_{} \phi\right)^2_{} +\mu_\Delta^2 \textrm{Tr}\left(\Delta^\dagger_{}\Delta\right) \nonumber\\
&&+ \lambda^{}_{\Delta} \left[\textrm{Tr}\left(\Delta^\dagger_{}\Delta\right)\right]^2_{} + \lambda'^{}_{\Delta} \textrm{Tr}\left(\Delta^\dagger_{}\Delta^\dagger_{}\right)  \textrm{Tr}\left(\Delta\Delta\right)  \nonumber\\
&&  +  \lambda''^{}_{\Delta}\textrm{Tr}\left(\Delta^\dagger_{}\Delta\Delta^\dagger_{}\Delta\right)+ \lambda^{}_{\phi\Delta}\phi^\dagger_{} \phi \textrm{Tr}\left(\Delta^\dagger_{}\Delta\right)  \nonumber\\
&&+ \lambda'^{}_{\phi\Delta}\phi^\dagger_{} \Delta\Delta^\dagger_{}\phi\,.
 \end{eqnarray}
Here we have forbidden the trilinear triplet-doublet coupling in the usual type-II seesaw model, i.e. 
\begin{eqnarray}
\label{trilinear}
V_{\phi\Delta}^{}&/\!\!\!\!\!\supset & \mu_{ \phi\Delta }^{}\left(\phi^T_{} i\tau_2^{} \Delta \phi + \textrm{H.c.} \right)\,.
 \end{eqnarray}
For this purpose, we impose a $Z_4^{}\times Z_4^{}$ discrete symmetry under which the Higgs triplet $\Delta$, the SM Higgs doublet $\phi$ and the SM fermions $F$ carry the charges as below,
\begin{eqnarray}
\label{z4z4}
\Delta(-,-)\,,~~\phi(+,+)\,,~~F(i,i)\,.
 \end{eqnarray}

The SM Higgs doublet $\phi$ spontaneously breaks the electroweak symmetry as usual, i.e.  
 \begin{eqnarray}
\phi = \left[\begin{array}{c} \frac{1}{\sqrt{2}} \left(v_\phi^{} + h_\phi^{}\right) \\
[2mm]
0\end{array}\right]~\textrm{with}~
v_\phi^{} = \sqrt{- \frac{\mu_\phi^2}{\lambda_\phi^{}}}\simeq 246\,\textrm{GeV}\,.
\end{eqnarray}
While the SM Higgs boson $h_\phi^{}$ has a mass \cite{pdg2018},
 \begin{eqnarray}
m_{h_\phi^{}}^2 = 2\lambda_\phi^{} v_\phi^2\simeq 125\,\textrm{GeV}~~\textrm{for}~~\lambda_\phi^{} \simeq 0.13\,,
\end{eqnarray} 
the neutral, singly-charged and doubly-charged components $(\delta^0_{},\delta^{\pm}_{},\delta^{\pm\pm}_{})$ of the Higgs triplet $\Delta$ acquire their masses as below, 
\begin{eqnarray}
\label{mass}
m_{\delta^{0}_{}}^{2} &=& \mu_\Delta^2 + \frac{1}{2}\lambda^{}_{\phi\Delta}v^2_{\phi} \,,
~~m_{\delta^{\pm}_{}}^{2} = m_{\delta^{0}_{}}^{2} +  \frac{1}{4}\lambda'^{}_{\phi\Delta}v_\phi^2  \,, \nonumber\\
m_{\delta^{\pm\pm}_{}}^{2} &=&  m_{\delta^{0}_{}}^{2} +\frac{1}{2} \lambda'^{}_{\phi\Delta}v_\phi^2  \,.
\end{eqnarray}
Clearly, the real and imaginary parts $\delta^0_{R,I}$ of the neutral component $\delta^0_{}$ are identical at this stage.

The above $\delta^0_R-\delta^0_I$ degeneracy can be broken by the following dimension-6 operator, 
\begin{eqnarray}
\label{eff}
 \mathcal{L}_{\textrm{eff}}^{} \supset \pm \frac{1}{\Lambda^2_{6}} \left[\left(\phi^T_{} i \tau_2^{} \Delta \phi \right)^2_{} + \textrm{H.c.}\right]\,.
\end{eqnarray}
The real part $\delta_R^{0}$ or the imaginary part $\delta_I^0$ now can be allowed to obtain a small quadratic term,
\begin{eqnarray}
\label{split1}
\!\! \!\!\!\!\!\! \!\!\!\!m_{\delta^{0}_{R}}^{2} \!&=&\!  m_{\delta^{0}_{}}^{2}  - \frac{v_\phi^4}{2\Lambda_6^2}  \ll  m_{\delta^0_I}^2 =  m_{\delta^{0}_{}}^{2} ~~\textrm{for}~~\frac{v_\phi^4}{2\Lambda_6^2} \simeq  m^2_{\delta^0_I}\,,\\
\!\! \!\!\!\!\!\! \!\!\!\! \textrm{or}\nonumber\\
\label{split2}
\!\! \!\!\!\!\!\!\!\!\!\! m_{\delta^0_R}^2 \! &=&\!  m_{\delta^{0}_{}}^{2} + \frac{v_\phi^4}{2\Lambda_6^2} \gg m_{\delta^{0}_{I}}^{2} = m_{\delta^{0}_{}}^{2}    ~~\textrm{for}~~\frac{v_\phi^4}{2\Lambda_6^2} \gg  m^2_{\delta^0_I}\,.
\end{eqnarray}
This means if its quadratic term is negative, i.e. $\mu_{\delta^{0}_{R(I)}}^{2} \equiv m_{\delta^{0}_{R(I)}}^{2}<0$, the $\delta_{R(I)}^{0}$ field can spontaneously develop a small VEV,
\begin{eqnarray}
\label{vdelta}
\delta_{R(I)}^{0}= v_\Delta^{}+  h_\Delta^{}~~ \textrm{with}~~v_\Delta^{} =  \sqrt{-\frac{\mu_{\delta^0_{R(I)}}^2}{\lambda^{}_\Delta + \lambda''^{}_\Delta }}\ll v_\phi^{}\,.
\end{eqnarray}
Accordingly, the Higgs boson $h_\Delta^{}$ can have a mass of the order of the VEV $v_\Delta^{}$, i.e.
\begin{eqnarray}
m_{h_\Delta^{}}^2=  2 (\lambda^{}_\Delta + \lambda''^{}_\Delta ) v_\Delta^2\,. 
\end{eqnarray}
It should be noted the Higgs bosons $h_\phi^{}$ and $h_\Delta^{}$ now are not the mass eigenstates. However, their mixing is suppressed by a factor $v_\Delta^{} / v_\phi^{}\ll 1$ so that it can be safely ignored.

\section{Neutrino mass and Hubble tension} 

Under the $Z_4^{} \times Z_4^{}$ symmetry (\ref{z4z4}), the SM Yukawa couplings are not be affected and are not shown for simplicity. Instead, we only give the Yukawa couplings involving the Higgs triplet $\Delta$, i.e. 
\begin{eqnarray}
\label{yukawa}
\mathcal{L}_Y^{} &\supset & - \frac{1}{2} f \bar{l}_L^c i \tau_2^{} \Delta l_L^{} + \textrm{H.c.}\,,
 \end{eqnarray}
with $l_L^{}$ being the SM lepton doublets,
\begin{eqnarray}
\begin{array}{c}l_L^{}(1,2,-\frac{1}{2})\end{array}\!&=&\left[\begin{array}{c}
\nu^{}_{L}\\
[1mm]
e^{}_{L}
\end{array}\right]\,.
\end{eqnarray}
Due to the above Yukawa interaction, the real part $\delta^0_{R}$ of the neutral component $\delta^0_{}$ from the Higgs triplet $\Delta$ is a scalar while the imaginary part $\delta_I^0$ is a pseudo-scalar.

As shown in Eq. (\ref{vdelta}), the Higgs triplet $\Delta$ can spontaneously develop a small VEV $v_\Delta^{}$. Therefore, the Yukawa couplings (\ref{yukawa}) can offer the neutrinos $\nu_L^{}$ a Majorana mass term, 
\begin{eqnarray}
\mathcal{L} &\supset & - \frac{1}{2} m_\nu^{} \bar{\nu}_L^c \nu_L^{} + \textrm{H.c.} ~~\textrm{with}\nonumber\\
&&m_\nu^{} = \frac{1}{\sqrt{2}}f v_\Delta^{} ~~\textrm{or}~~m_\nu^{} = \frac{i}{\sqrt{2}}f v_\Delta^{} \,. 
 \end{eqnarray}
We would like to emphasize that the above neutrino mass generation is different from the usual type-II seesaw \cite{mw1980,sv1980,cl1980,lsw1981,ms1981} where the Higgs triplet should have a trilinear coupling with the SM Higgs doublet. See Eq. (\ref{trilinear}). Therefore, our split Higgs triplet $\Delta$ can not mediate the well-known Weinberg dimension-5 operator \cite{weinberg1979}, i.e.
\begin{eqnarray}
\mathcal{L}_{\textrm{eff}}^{}  &/\!\!\!\!\!\supset & -\frac{1}{\Lambda_5^{}} \bar{l}_{L}^{} \phi \phi^T_{} l_L^c + \textrm{H.c.}\,.
 \end{eqnarray}
Indeed the Weinberg dimension-5 operator and the other high-dimensional lepton-number-violating operators are all protected by the $Z_4^{} \times Z_4^{}$ symmetry (\ref{z4z4}).

The Higgs boson $h_\Delta^{}$ may not be allowed below the MeV scale by the BBN. The VEV $v_\Delta^{}$ thus may be larger than the MeV scale. In this case, the Yukawa couplings $f$ should be of the order of $\mathcal{O}(10^{-7}_{})$ to make the neutrino mass $m_\nu^{}$ of the order of $\mathcal{O}(0.1\,\textrm{eV})$ \cite{pdg2018}. Such small Yukawa couplings can be understood by resorting to some vector-like lepton doublets which couple to the split Higgs triplet $\Delta$ and mix with the SM lepton doublets $l_L^{}$. The details can be found in a recent work \cite{gu2020}. 

In the other case that the light scalar $\delta^0_R$ or the light pseudo-scalar $\delta^0_I$ has a positive quadratic term and does not develop a VEV, the Yukawa couplings (\ref{yukawa}) will not be constrained by the neutrino mass. This means we can have the flexibility to choose the structure and the size of these Yukawa couplings. For example, the Higgs triplet $\Delta$ can mostly even exclusively couple to the tau neutrino. Especially, the tau-neutrino self-interaction mediated by the light scalar $\delta^0_R$ or the light pseudo-scalar $\delta^0_I$ can fall in a so-called moderately interacting regimes \cite{bkkm2019},
\begin{eqnarray}
\label{mir}
\mathcal{L}_{\textrm{eff}}^{} &\supset& \pm G_{\textrm{eff}}^{}  \bar{\nu}_{L\tau}^c \nu_{L\tau}^{} \bar{\nu}_{L\tau}^{} \nu_{L\tau}^c~~\textrm{with}\nonumber\\
&&G_{\textrm{eff}}^{}   = \frac{f_{\tau\tau}^2}{2m_{\delta^0_{R(I)}}^2}=\left(89^{+171}_{-61} \,
\textrm{MeV}\right)^{-2}_{}\,.
 \end{eqnarray}
The Hubble tension thus can be resolved while the other experimental constraints can be satisfied \cite{bkkm2019}. From Eq. (\ref{mir}), we can put an upper bound on the mass of the light scalar $\delta_R^0$ or the light pseudo-scalar $\delta^0_I$, i.e.
\begin{eqnarray}
m_{\delta^0_{R(I)}}^{}<223^{+429}_{-153} \,
\textrm{MeV} ~~\textrm{for}~~f_{\tau\tau}^{}< \sqrt{4\pi} \,.
 \end{eqnarray}

In order to simultaneously generate the neutrino mass and reconcile the Hubble tension, we can introduce two split Higgs triplets. However, it may be difficult to distinguish these two split Higgs triplets by certain symmetries. Since the Majorana nature of neutrinos is just a theoretical assumption and has not been confirmed by any experiments, we can consider some Dirac seesaw \cite{rw1983,rs1984,mp2002,gh2006,gu2016} for light Dirac neutrinos while the split Higgs triplet can focus on the Hubble tension.

\section{Constraints and implications}

Besides the perturbation condition, the scalar potential (\ref{potential}) should be bounded from below in all the directions of the field space. We hence require
\begin{eqnarray}
\!\!\!\!\!\!\!\!\!\!\!\!\!\!\!\!&&0\leq \lambda_{\Delta}^{}+\lambda''^{}_{\Delta}\,,  \lambda_{\Delta}^{}+\lambda'^{}_{\Delta}+\frac{1}{2}\lambda''^{}_{\Delta} <4\pi\,,\nonumber\\
\!\!\!\!\!\!\!\!\!\!\!\!\!\!\!\!&&-2\sqrt{\lambda_\phi^{}\left(\lambda^{}_\Delta + \lambda''^{}_\Delta\right)} \leq \lambda^{}_{\phi\Delta} \,,  \lambda^{}_{\phi\Delta} +\lambda'^{}_{\phi\Delta} <4\pi \,,\nonumber\\
\!\!\!\!\!\!\!\!\!\!\!\!\!\!\!\!&&-2\sqrt{\lambda_\phi^{}\left(\lambda^{}_\Delta + \lambda'^{}_\Delta +\frac{1}{2}\lambda''^{}_\Delta\right)} \leq \lambda^{}_{\phi\Delta} +\frac{1}{2}\lambda'^{}_{\phi\Delta}<4\pi \,. \end{eqnarray}
The coupling $\lambda_{\phi\Delta}^{}$ should be also constrained by the invisible decay of the SM Higgs boson $h_\phi^{}$ into the light scalar $\delta^0_{R}$ or the light pseudo-scalar $\delta^0_{I}$, i.e.
\begin{eqnarray}
\Gamma(h_\phi^{} \rightarrow \delta^0_{R,I}\delta^0_{R,I}) = \frac{\lambda_{\phi\Delta}^2 }{32\pi} \frac{v_\phi^2}{m_{h_\phi^{}}^{}}\,.
 \end{eqnarray}
We can take the coupling $\lambda_{\phi\Delta}^{}$ small enough to satisfy the limit on the $h_\phi^{}$ invisible branching fraction \cite{pdg2018}. By ignoring the coupling $\lambda_{\phi\Delta}^{}$, we can put an upper bound on the charged scalar masses in Eq. (\ref{mass}), i.e.
\begin{eqnarray}
m_{\delta^{\pm}_{}}^{2} &<& m_{\delta^{0}_{I}}^{2} + \left(436\,\textrm{GeV}\right)^2_{}  \,, \nonumber\\
 m_{\delta^{\pm\pm}_{}}^{2} &<&  m_{\delta^{0}_{I}}^{2} + \left(617\,\textrm{GeV}\right)^2_{} ~~\textrm{for}~~ \lambda'^{}_{\phi\Delta} < 4\pi\,.
\end{eqnarray}
This means even if the pseudo-scalar $\delta^0_{I}$ is very light, the charged scalars $\delta^{\pm}_{}$ and $\delta^{\pm\pm}_{}$ can be heavy enough to test at the LHC \cite{fhhlw2008,ddry2019}.

When the scalar $\delta^0_R$ (the pseudo-scalar $\delta^0_I$) is very light, the pseudo-scalar $\delta^0_I$ (the scalar $\delta^0_R$) should be heavy enough to suppress the invisible decay width of the $Z$ boson, i.e.
\begin{eqnarray}
\Gamma(Z \rightarrow \delta^0_{R}\delta^0_{I}) =\frac{G_F^{} m_Z^3}{2\sqrt{2}\pi}\left(1-\frac{m_{\delta^0_{R,I}}^2}{m_Z^2}\right)^{3}_{}\,.
 \end{eqnarray}
We thus can simply take the scalar $\delta^0_R$ or the pseudo-scalar $\delta^0_I$ heavier than the $Z$ boson to match the experimental uncertainty in the invisible $Z$ width \cite{pdg2018}. Consequently we can give an upper bound on the effective cutoff $\Lambda_6^{}$ in the dimension-6 operator (\ref{eff}), i.e.
\begin{eqnarray}
m_{\delta^0_{R,I}}^2 \simeq \frac{v_\phi^4}{2\Lambda_6^2} > \left(91\,\textrm{GeV}\right)^2_{}~~\Rightarrow ~~\Lambda_6^{}< 470\,\textrm{GeV}\,,
\end{eqnarray}
from Eq. (\ref{split1}) or Eq. (\ref{split2}).

In the usual type-II seesaw model where the scalar $\delta^0_R$ and the pseudo-scalar $\delta^0_I$ are both heavy, the following decay channels should be most important for the Higgs triplet searches \cite{fhhlw2008,ddry2019},
\begin{eqnarray}
\label{search1}
&&\delta^{\pm}_{}\rightarrow W^{\pm}_{}Z\,,~~\delta^{+}_{}\rightarrow e^{+}_{i} \nu_{Lj}^{}\,,~~\delta^{-}_{}\rightarrow e^{-}_{i} \bar{\nu}_{Lj}\,,\nonumber\\
&&\delta^{\pm\pm}_{}\rightarrow W^{\pm}_{} W^{\pm}_{}\,,~~\delta^{\pm\pm}_{}\rightarrow e^{\pm}_{i} e^{\pm}_{j}\,.
\end{eqnarray}
Now the light scalar $\delta^0_R$ or the light pseudo-scalar $\delta^0_I$ can have a different phenomenology, i.e.
\begin{eqnarray}
\label{search2}
&&\delta^{0}_{R(I)}\rightarrow Z \delta^{0}_{I(R)}\,,~~\delta^{\pm}_{}\rightarrow W^{\pm}_{}\delta^{0}_{R,I}\,,\nonumber\\
&&~\delta^{\pm\pm}_{}\rightarrow W^{\pm}_{} W^{\pm}_{} \delta^{0}_{R,I}\,.
\end{eqnarray} 
This can be understood by the fact that the decays (\ref{search1}) rather than the decays (\ref{search2}) are suppressed by the small VEV $v_\Delta^{}$ or the small Yukawa couplings $f$. Of course, the Yukawa couplings $f$ can be sizable to dominate the decaying processes in Eq. (\ref{search1}) and (\ref{search2}) when the split Higgs triplet does not develop a VEV for the Majorana neutrino mass generation.

\section{Renormalizable models}

\begin{figure*}
\vspace{6cm} \epsfig{file=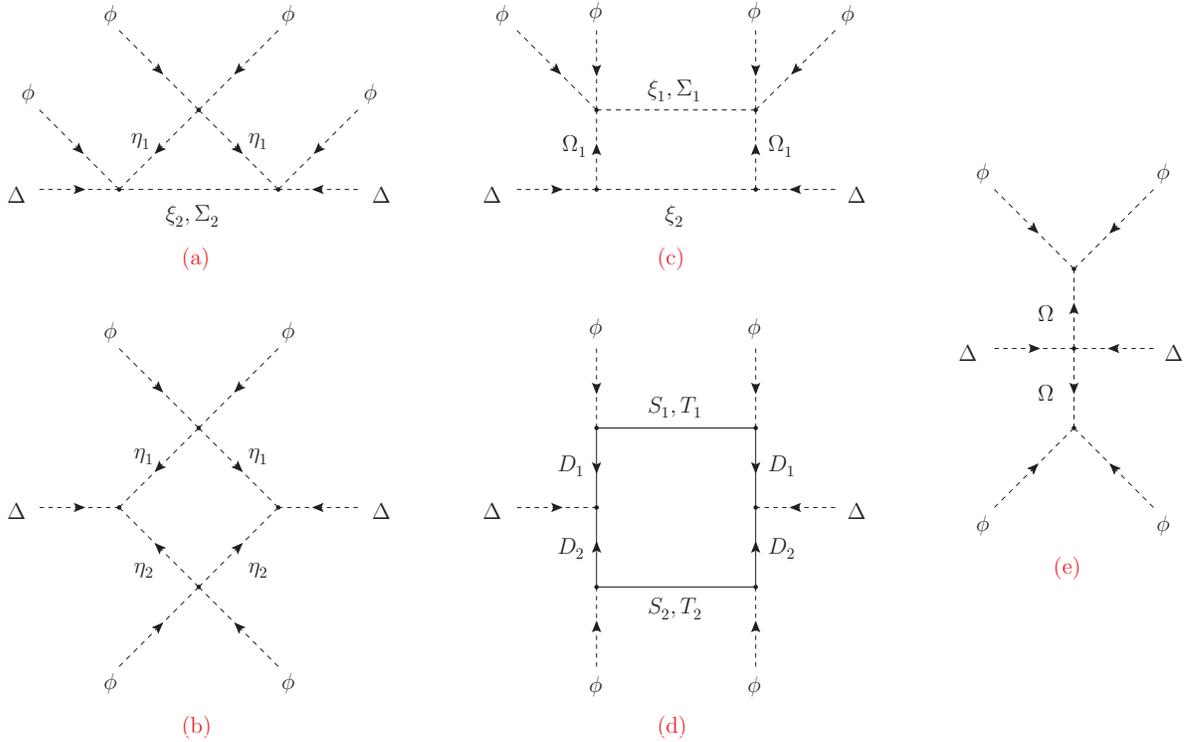, bbllx=10cm, bblly=6.0cm,
bburx=20cm, bbury=16cm, width=6cm, height=6cm, angle=0,
clip=0} \vspace{-1cm} \caption{\label{dim-6} Some renormalizable models for split Higgs triplet. }
\end{figure*}

So far our discussions on the split Higgs triplet are based on the effective theory. It should be more attractive if this scenario arises from a renormalizable theory. For this purpose, we construct five possible models. The relevant diagrams are shown in Fig. \ref{dim-6}. In the one-loop diagrams (a-d), we introduce the scalars $\xi_{1,2}^{}(1,1,0)$, $\eta_{1,2}^{}(1,2,-1/2)$, $\Sigma_{1,2}^{}(1,3,0)$ and $\Omega_{1}^{}(1,3,+1)$ as well as the Majorana singlets $S_{1,2}^{}=S_{R1,2}^{}(1,1,0)+\left(S_{R1,2}^{}(1,1,0)\right)^{c}_{}$, the vector-like doublets $D_{1,2}^{}=D_{L1,2}^{}(1,2,-1/2)+\left(D_{L1,2}^{}(1,2,+1/2)\right)^c_{}$ and the Majorana triplets $T_{1,2}^{}=T_{L1,2}^{}(1,3,0)+\left(T_{L1,2}^{}(1,3,0)\right)^c_{}$. For the $Z_4^{}\times Z_4^{}$ symmetry (\ref{z4z4}), the fields with subscript $"1"$ and $"2"$ carry the parity $(-,+)$ and $(+,-)$ respectively. In the tree diagram (e), the scalar $\Omega(1,3,+1)$ denotes one of the Georgi-Machacek Higgs triplets.

As an example, we demonstrate a simple model in the diagram (a). Specifically we consider an inert real singlet scalar and an inert Higgs doublet,  
\begin{eqnarray}
\xi\,,~~\eta=\left[\begin{array}{c}
\eta^{0}_{}=\frac{1}{\sqrt{2}}\left(\eta^0_R+ i \eta^0_I\right)\\
[1mm]
\eta^{-}_{}
\end{array}\right]\,,
 \end{eqnarray}
where the subscripts of the fields in the diagram have been omitted for simplicity. The $\xi-\eta$ inert scalars have the following potential,
\begin{eqnarray}
V_{\xi\eta}^{}&=& \frac{1}{2}\mu_\xi^2 \xi^2_{} +  \frac{1}{4}\lambda_\xi^{}\xi^4_{} + \mu_\eta^2 \eta^\dagger_{} \eta+  \lambda_\eta^{}\left(\eta^\dagger_{} \eta\right)^2_{} \nonumber\\
&&+ \frac{1}{2}\lambda_{\xi\eta}^{} \xi^2_{} \eta^\dagger_{}\eta\,,
\end{eqnarray}
as well as the interactions with the $\phi-\Delta$ Higgs scalars, 
\begin{eqnarray}
V_{\xi\eta-\phi\Delta}^{}&=&\frac{1}{2} \kappa_{1}^{} \xi^2_{}   \phi^\dagger_{} \phi  + \frac{1}{2}\kappa_{2}^{} \xi^2_{}  \textrm{Tr}\left(\Delta^\dagger_{}\Delta\right)+ \kappa_{3}^{} \eta^\dagger_{}\eta \phi^\dagger_{} \phi   
\nonumber\\
&&+ \kappa_{4}^{} \eta^\dagger_{}\phi  \phi^\dagger_{}\eta+ \frac{1}{2}\kappa_{5}^{}\left[\left(\eta^\dagger_{}\phi\right)^2_{}+\textrm{H.c.}\right] \nonumber\\
&&+\kappa_{6}^{}\eta^\dagger_{} \eta\textrm{Tr}\left(\Delta^\dagger_{}\Delta\right)  + \kappa_{7}^{}\eta^\dagger_{} \Delta^\dagger_{}\Delta\eta \nonumber\\
&&+ \kappa_{8}^{}\eta^\dagger_{} \Delta\Delta^\dagger_{}\eta +  \kappa_{9}^{} \xi  \left(\eta^T_{}i\tau_2^{} \Delta \phi  + \textrm{H.c.}\right)\,.
 \end{eqnarray}

After the electroweak symmetry breaking, the inert scalars can obtain their physical masses, 
\begin{eqnarray}
m_{\eta^0_I}^2 &=&m_{\eta^{\pm}_{}}^2 + \frac{1}{2}\kappa_4^{}v_\phi^2 \,,~~m_{\eta^0_R}^2 = m_{\eta^{0}_{I}}^2 + \frac{1}{2} \kappa_5^{}v_\phi^2 \,,\nonumber\\
m_{\eta^\pm_{}}^2 &=&  m_{\eta^{0}_{I}}^2- \frac{1}{2}\kappa_{4}^{}v_\phi^2\,,~~m_\xi^2 = \mu_\xi^2 + \frac{1}{2}\kappa_{1}^{}v_\phi^2 \,.
\end{eqnarray} 
Their contributions to the quadratic term of the light scalar $\delta_R^{0}$ then can be well computed by 
\begin{eqnarray}
\frac{v_\phi^4}{2\Lambda_6^2} &=& \frac{\kappa_9^2 v_\phi^2 }{32\pi^2_{}}\left[\frac{m_{\eta^0_R}^2}{m_{\eta^0_R}^2-m_\xi^2} \ln\left(\frac{m_{\eta^0_R}^2}{m_\xi^2}\right)  \right.\nonumber\\
&&\left.- \frac{m_{\eta^0_I}^2}{m_{\eta^0_I}^2-m_\xi^2} \ln\left(\frac{m_{\eta^0_I}^2}{m_\xi^2}\right) \right]\,.
\end{eqnarray} 
The effective cutoff $\Lambda_6^{}$ can be simply read in some limiting cases \cite{foot1}, 
\begin{eqnarray}
\label{assumption}
\Lambda_6^2 \!&\simeq & \!\frac{16\pi^2_{}}{\kappa_5^{} \kappa_9^2 } m_{\eta^0_I}^2 = \!\left(36\,\textrm{GeV}\right)^2_{}\!\left(\frac{4\pi}{\kappa_5^{}}\right)\!\left(\frac{4\pi}{ \kappa_9^{} }\right)^2_{}\!\left(\frac{ m_{\eta^0_I}^{}}{125\,\textrm{GeV}}\right)^2_{}\nonumber\\
\!&&\!\textrm{for} ~~ m_{\eta^0_R}^2 \gg m_{\eta^0_I}^2 \gg m_\xi^2\,.
\end{eqnarray} 

The mixing between the $\xi$ and $\eta^{0}_{R,I}$ scalars is highly suppressed by the small VEV $v_\Delta^{}$. So, the $\xi$ and $\eta^{0}_{R,I}$ scalars can well approximate to the mass eigenstates. We then take $\xi$ to be lighter than $\eta^0_{R,I}$ for the simplification (\ref{assumption}). This means a stable $\xi$ for a residual $Z_4^{}\times Z_4^{} \rightarrow Z_2^{}$ symmetry. In the presence of the light scalar $\delta^0_R$ or the light pseudo-scalar $\delta^{0}_{I}$, the stable scalar $\xi$ can serve as a dark matter particle through the annihilation $\xi\xi \rightarrow \delta^{0}_{R(I)}\delta^{0}_{R(I)}$.

\section{Summary}

In this paper we have proposed a new idea of split Higgs triplet \cite{foot2} with rich observable phenomena. When the split Higgs triplet spontaneously develops a small VEV, the neutrinos can obtain their Majorana masses without causing any high-dimensional lepton number violations including the well-known Weinberg dimension-5 operator. Alternatively the split Higgs triplet can mediate a non-standard neutrino self-interaction to resolve the reported Hubble tension. We demonstrate such split Higgs triplet in an effective theory and then realize it in various renormalizable models \cite{foot3} with the Georgi-Machacek Higgs triplets or some dark matter fields.

\textbf{Acknowledgement}: This work was supported in part by the National Natural Science Foundation of China under Grant No. 11675100 and in part by the Fundamental Research Funds for the Central Universities.


\begin{thebibliography}{99}
\bibitem{foot1}
We may introduce more inert real singlet scalars and/or more inert Higgs doublets to lower the effective cutoff.
\bibitem{foot2}
Similarly we can split the neutral and charged components of a real Higgs triplet without hypercharge.
\bibitem{foot3}
An inert real singlet scalar and the scalar component of an inert Higgs triplet can mix with each other after the electroweak symmetry breaking. These two scalars with sizable mixing can leave a light mass eigenstate and a heavy mass eigenstate.
\end{thebibliography}
\end{document}